# *Efficient infrared sunlight absorbers based on gold-covered, inverted silicon pyramid arrays*

Jinhui Hu[a], Luis A. Pérez[a,*], Juan Luis Garcia-Pomar[b], Agustín Mihi[a], Miquel Garriga[a], M. Isabel Alonso[a], Alejandro R. Goñi[a,c,**]

[*]lperez@icmab.es, [**]goni@icmab.es

[a]*Institut de Ciència de Materials de Barcelona, ICMAB-CSIC, Campus UAB, 08193 Bellaterra, Spain*

[b]*INL-International Iberian Nanotechnology Laboratory, Av. Mestre José Veiga s/n, 4715-330 Braga, Portugal*

[c]*ICREA, Passeig Lluís Companys 23, 08010 Barcelona, Spain*

Abstract:

The transparency of silicon in the infrared region enables the design of nano/microstructures for implementation in devices to harvest the infrared (IR) part of the solar spectrum. Herein we report a strategy that uses arrays of inverted silicon pyramids covered with a thin gold film, which exhibit substantial light absorption in the infrared spectral range (below the gap of Si). The absorption stems from the resonant excitation at infrared wavelengths of surface-plasmon polaritons at the metal/dielectric interface mainly by tuning size and separation of the inverted pyramids. The array-parameters optimization proceeded by iteration of the calculation and measurement of the infrared response using finite difference time-domain simulations and Fourier-transform IR spectroscopy, respectively. We show two fabrication routes for this kind of metal/silicon metamaterials either by photolithography or scalable nanoimprint techniques for a seamless integration in optoelectronic fabrication processes.

Introduction:

Besides dominating electronics industry, silicon also finds wide applications in photovoltaics, primarily due to its cost-effective and experienced manufacturing process[1–3]. However, as in any solar cell, the exploitable spectral range of the sunlight is limited by the band gap of the active photovoltaic material, which in the case of silicon is 1.1 eV, leading to an absorption edge roughly at 1100 nm. Thus, nearly half of the solar spectrum remains unabsorbed, namely, the near infrared (NIR) part[4]. To alleviate this situation, a common alternative to photoelectric conversion relies on the mechanism of plasmon-enhanced internal photoemission of hot electrons at metal/semiconductor hetero-structures. NIR photons with energies below the gap of the semiconductor are absorbed by the metal, launching surface plasmons, which, upon relaxation, produce a photocurrent of hot carriers into the semiconductor[5-8]. For an overview of the different applications of plasmon-induced hot electrons, including solar energy harvesting, we refer to several recent reviews and references therein[9-12].

The excellent concentration and light manipulation properties of surface plasmons can be further enhanced by nanostructuring of the metal/semiconductor system[6,13]. In this respect, silicon offers a fairly simple opportunity for its texturing in the form of both inverted and upright square pyramids, due to the strong anisotropy in the etching speed along different crystallographic planes, when KOH is used as etcher. This kind of nanostructuration of the metal-silicon interfaces has already been used to enhance the performance of Si solar cells[14-16]. Furthermore, inverted Si pyramid arrays has been successfully implemented as surface enhanced Raman scattering (SERS) substrates, either directly[17] or by serving as template for the preparation of upright bulk metallic pyramids[18] or pyramidal arrays made of perfectly piled gold nanoparticles[19]. For the purpose of this work, however, the ansatz of implementing both types of pyramids, covered with different metals (Au, Al and Cu), in internal-photoemission detectors is of great interest. In fact, promising responsivity figures of merit have been already reported for the NIR wavelength range (ca. 1 – 2.7 μm) [20-23].

In this work, we propose a strategy for harvesting NIR sunlight based on the fabrication of silicon inverted-pyramid arrays covered with thin Au films. The textured structures are obtained by scalable soft nanoimprint lithography, subsequent wet KOH etching and thermal Au deposition. A germanium layer deposited onto the Si wafer by molecular-beam epitaxy (MBE) serves as hard mask in the fabrication process of nano/micro-scale inverted pyramids due to its easy dissolution in hydrogen peroxide ($H_2O_2$) but high resilience against KOH. By illumination from the semiconductor side with NIR light (wavelengths between 1 and 2.5 microns), surface-plasmon polaritons can be excited at the Au/Si nanostructured interface, leading to well-defined absorption bands in the NIR transmission spectra and the converse features in reflectance. The calculated near-field distributions for the observed surface-plasmon resonances indicate the existence of hot spots along the contour of the pyramids, where the electromagnetic field intensity becomes strongly enhanced (up to two orders of magnitude, depending on pyramid size) with field vector components perpendicular to the pyramid facets. This allows us to foresee great potential for Au/Si pyramidal arrays for the generation of hot-carrier photocurrents by internal photoemission for energy harvesting applications or photodetectors with improved sensitivity in the infrared.

Methods:

*Fabrication of inverted nano-pyramids*: Arrays of inverted Si pyramids with sizes roughly around 500 nm were fabricated by nanoimprint lithography following a similar method as published elsewhere[24]. Different key steps of the fabrication process are depicted in Fig. 1 together with the resulting nanostructures, as shown by the corresponding secondary electron micrographs (SEM). First a 150-nm thick Ge layer was deposited by MBE on a one-side polished silicon wafer with a thickness of 500 μm[25]. Germanium was chosen as hard mask because it can be easily dissolved by $H_2O_2$ with controllable etching rate at room temperature[26], as compared with conventional toxic and risky hydrogen fluoride (HF), the etchant used to remove silicon dioxide, for example. Furthermore, the thin Ge layer is robust in hot alkaline KOH solution used for etching the inverted pyramids. On top of the Ge, a 600-nm thick SU8 photoresist (Microchem) layer was spin-coated at 2000 rpm for 10 s. Subsequently, different stamps of polydimethylsiloxane (PDMS) were used to imprint a square array of holes into the SU8 photoresist. The PDMS moulds consisted in arrays of pillars with a diameter of 300 nm, a height of 350 nm but a different lattice parameter (LP) of 400, 500 and 600 nm; the pattern covering an area of 16 mm$^2$. The nanoimprint step took place on a hot plate at 90 $^o$C, immediately after the coating with the photoresist. For this purpose, the patterned PDMS moulds were gently pressed against the substrate for 30 s and then left to cool down to below the glass transition temperature of the resist. The negative pattern of the PDMS mould was obtained on the SU8 layer by gently stripping off the master, while it cools down to room temperature. The patterned SU8 layer was cured under UV light for 10 min and hard-baked at 160 $^o$C on a hot plate for 30 min. The remaining resist and Ge in the holes were removed by reactive ion etching (RIE) using a PlasmaPro Cobra 100 system (Oxford Instruments). The residual resist was etched by $O_2$ plasma at 50 sccm, 12 mTorr and a high field power of 50 W for 2 min, whereas for the Ge layer a mixture of $O_2/SF_6$ at 5/30 sccm, 15 mTorr and a high field power of 75 W for 2 min was employed for RIE. The inverted pyramids are then engraved on the silicon by anisotropic wet etching using a 33 wt% KOH solution at 85 $^o$C for 30 seconds, then rinsed with water quickly. The residual Ge mask is then removed using 30% $H_2O_2$ for 5 min, rinsed with water and dry blown with $N_2$. Finally, a 50 nm thick Au layer was deposited on the surface of the inverted pyramids by thermal evaporation. As shown by the SEM pictures in Figs. 1e and 1f, we were able to fabricate high-quality large-area arrays of perfectly ordered and identical inverted Si/Au pyramids.

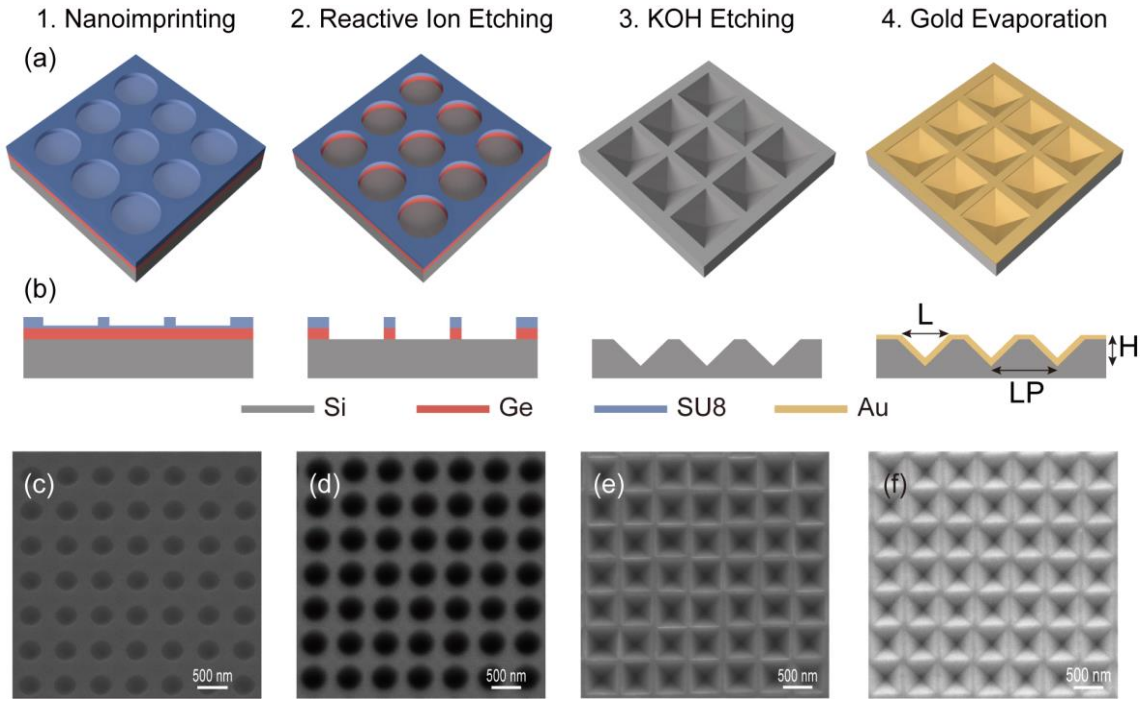

**Figure 1:** Fabrication of the arrays of nano-scale inverted pyramids covered with Au film. **(a)** Schematic of the main steps of the fabrication process. **(b)** Corresponding cross-section sketches for each process step (1-4). **(c-f)** Top view scanning electrons microscopy (SEM) images corresponding to process steps (1-4).

*Fabrication of inverted micro-pyramids*: Larger pyramids with sizes ranging from 3 to 5 microns were fabricated by conventional photolithography for practical reasons. A sketch of the different steps of the fabrication procedure is displayed in Fig. S1 of the ESI. For the fabrication of microns-size pyramids a 1200 nm-thick Shipley resist was deposited on Si/Ge heterostructure at 5000 rpm for 25 s and baked on a hotplate at 95 °C for 1 min. Then, a pattern of holes with the desired diameter and pitch is printed on the resist with a ML3 micro-writer (Durham Magneto Optics) under exposition with a 385 nm laser, subsequently developed for 45 s and rinsed with water. A wet $H_2O_2$ etching was carried out to remove rest of the Ge in the holes for 4 min at room temperature. An anisotropic wet etching in 33wt% KOH solution at 80 °C for 4 min produced again the inverted pyramids in the exposed Si and then the etched sample was rinsed with water and blown dry with $N_2$. Finally, a 50 nm thick Au film was deposited by thermal evaporation on the surface of inverted pyramid.

*Numerical Simulation*: The far-field optical response (reflection and transmission) as well as the near-field distributions of the Si/Au inverted-pyramid structures were simulated using commercial software (Lumerical) based on finite-difference time domain (FDTD) method. The numerical simulations were performed for linearly polarized light at normal incidence from the silicon substrate side and in the NIR wavelength range (1000 – 2500 nm). Literature data were employed to account for the wavelength-dependent refractive index and extinction coefficient of gold[27] and silicon[28]. The geometry of the simulated structures was obtained from the SEM micrographs of the actual samples. Figure 2 shows sketches of the geometry employed for the FDTD simulations regarding direction of incidence of the light, given by the wave vector **k**, its linear polarization corresponding to the electric field vector **E** with respect to the plane of the monitor (purple indicated by the planes) used for the calculation of the near fields. Calculations were performed for two different planar monitors passing through the centre of a pyramid either bisecting two facets (0º monitor) or along two diagonals (45º monitor). In this configuration, the incident light was either S- (perpendicular) or P-polarized (parallel) to the monitor plane, as indicated by the sketches in Fig. 2.

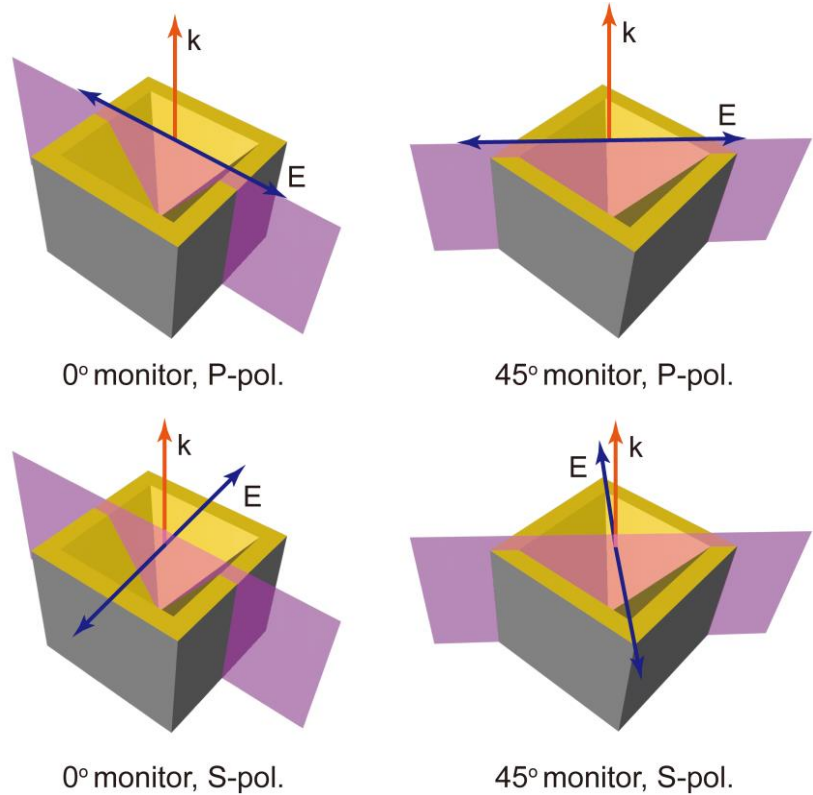

**Figure 2:** Sketches of the geometry used for the FDTD calculations including the cross-section monitor planes passing through the centre of the inverted pyramids either bisecting two facets (0º monitor) or along two diagonals (45º monitor). The light with wave vector k is either S-polarized or P-polarized, i.e. perpendicular or parallel to the monitor plane, respectively.

A single inverted pyramid was set as the unit cell for the simulations (see Fig. S2 of the ESI) using periodic boundary conditions (PBC) along the x and y axes and perfectly matched layer conditions along the z-axis. The pyramid and the surrounding medium were divided into a mesh for computing the finite differences. A variable mesh-size method with a minimum element size of 4 nm was adopted to save computational time and resources. For microscale pyramids, hence, the smallest mesh size was 20 nm.

*Scanning electron microscopy*: SEM images were acquired with a Quanta FEI 200 FEG-ESEM microscope at an acceleration voltage of 15 kV, equipped with a field-emission gun source. All images were obtained in high vacuum and detecting the secondary electron signal.

*Optical characterization*: Infrared transmission and reflectance spectra were recorded with a Fourier-transform infrared (FTIR) spectrometer attached to an optical microscope (Vertex 70 and Hyperion, Bruker). A ×4 magnification objective with 0.10 numerical aperture provides a 2×2 mm field of vision. A gold mirror with high reflectivity (≥98%) was adopted as reference for the reflectance spectra, whereas the transmission spectra were normalized by the intensity of the incident beam.

Results and discussion:

Figure 3 shows SEM images of two representative arrays of inverted pyramids after the deposition of the thin gold film. The notation for the morphological parameters is shown in the sketches of Fig.1, which includes the pyramid base length L and height H as well as the lattice parameter LP of the array. Figure 3 thus displays the results for an array of small (L=470 nm) and large (L=4.1 μm) inverted pyramids. In both cases, the arrays exhibit extremely high regularity in shape and size of the pyramids. The insets to both panels of Fig. 3 show a magnified view of a single pyramid, indicating again in both cases that the inverted pyramids are fully developed after KOH etching and that the surface of the deposited gold is continuous and smooth. We note that the regularity and high quality of the produced arrays typically extend over the entire pattern area up to 10x10 mm$^2$.

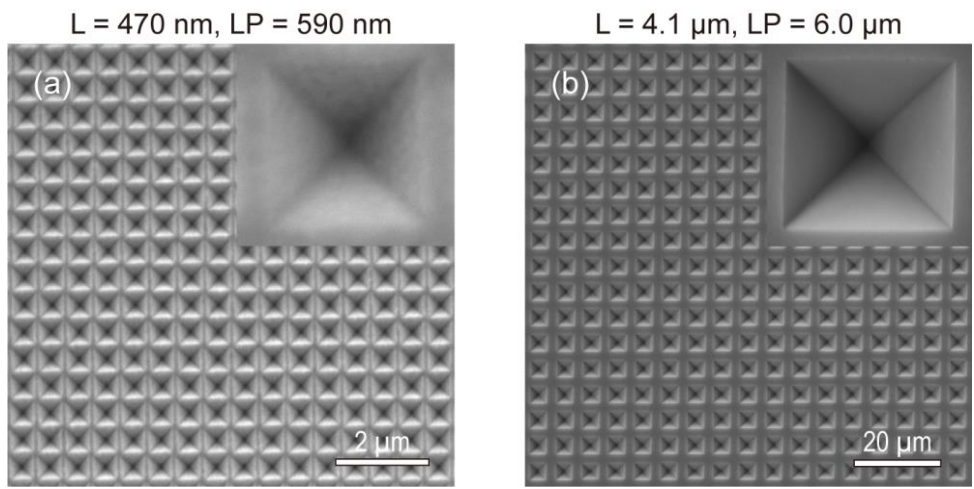

**Figure 3:** Inverted pyramid arrays covered with a 50-nm thick Au film. **(a)** SEM top-view image of representative small inverted pyramids with average L=470(20) nm and LP=590(20) nm. **(b)** SEM top-view image of representative large inverted pyramids with average L=4.1(2) μm and LP=6.0(2) μm. **Insets:** high resolution images of a single pyramid. Numbers in parenthesis indicate the experimental uncertainty.

Before discussing the optical properties of the inverted-pyramids, we show the results of a pre-screening of some array parameters, performed for sub-micron size pyramids in view of the computational cost of the FDTD calculations. For a given pyramid size the lattice parameter LP or pitch is crucial for the tuning of the optical response of the array. Figure 4 shows contour plots of the normalized absorption, calculated as a function of the incident wavelength and the lattice parameter LP, for an array of inverted pyramids with L=500 nm covered with a 50-nm thick Au layer. It is instructive to analyse first the results of Fig. 4 in more detail. For a given lattice parameter the normalized absorption exhibits a series of maxima which correspond to the different plasmon polaritons exited by the incident (white) light. From the evolution of the different maxima with increasing LP we can clearly distinguish between two types of plasmon polaritons[29]. The ones exhibiting a clear dispersion are called Bragg modes and correspond to propagating extended modes with strong photonic character. In contrast, the dispersionless modes (vertical lines in the LP versus wavelength plot) are called Mie modes and correspond to strongly localized modes. From the plot on Fig. 4 it is clear that for any lattice parameter there is always certain degree of mixing between Mie and Bragg modes. As shown below, this has an impact on the plasmonic/photonic character of the near field associated with the different polariton modes.

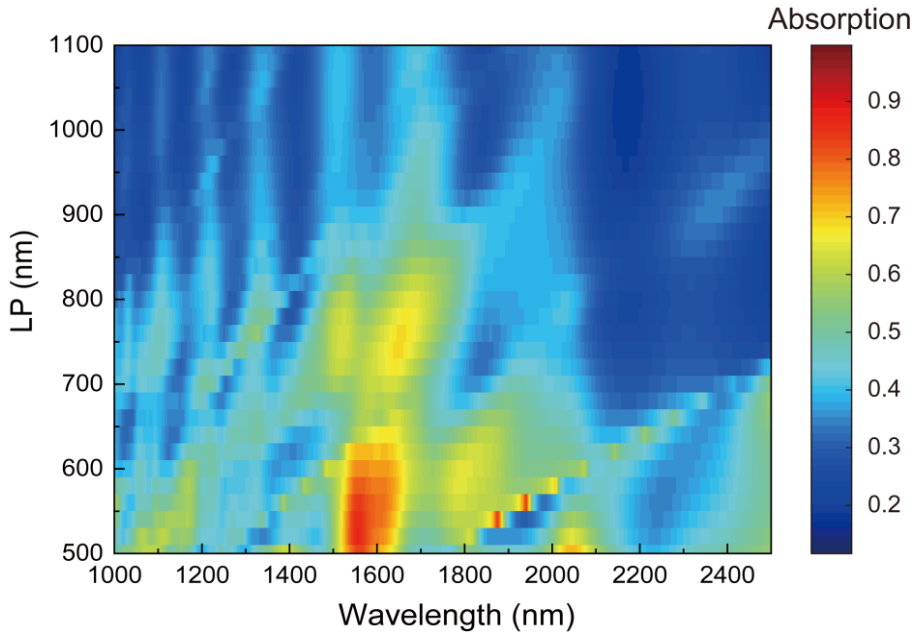

**Figure 4:** Contour plots of the normalized absorption as a function of the incident wavelength and the lattice parameter LP for an array of inverted pyramids with L=500 nm covered with 50 nm of gold.

Regarding the dependence of the absorption on the array lattice parameter LP, we infer from the plot of Fig. 4 that the resonances quickly loose intensity with increasing LP values. This is a consequence of the obvious fact that with increasing array pitch but constant pyramid size the ratio between patterned/flat areas also decreases. We have thereafter fabricated arrays with lattice parameters only 10% to 20% larger than the pyramid base size. In Fig. S3 of the ESI, we also show the pre-screening of the gold layer thickness. The result is again straightforward: With increasing Au thickness light absorption increases rapidly from zero to a saturation value, reached above a thickness of ca. 30 nm. Consequently, we deposited routinely 50 nm of gold, so as to obtain smooth, strongly-absorbing Au layers.

Figure 5 shows reflectance spectra measured in the NIR range (1000 to 2500 nm) of a small-pyramid array (L=470 nm, LP=590 nm) before and after its coating with Au. The light, incident from the substrate side (indicated by its wave vector **k**), was either unpolarised or linearly polarized perpendicular to two pyramid facets (denoted as 0º) or along two pyramid diagonals (denoted as 45º), as depicted in Fig. 2. For bare inverted pyramids without Au film, the overall reflectance for unpolarised light (black curve) is roughly featureless, ranging from 37~48%. In contrast, the thin gold layer has large impact of the optical response of the array (red, blue, green curves). From 1000 to 2000 nm the reflectance baseline decreases to less than 30%, but exhibiting a prominent broad peak at about 1600 nm. Above 2000 nm, the reflectance suddenly jumps to a constant value of ca. 56%. This edge-like profile arises from an optical phenomenon known as Rayleigh anomaly (RA), which is associated with light diffracted parallel to the surface of the regular array, acting as a grating[30]. The lowest order RA occurs at a wavelength given by:

$$\lambda_{RA} = n_{Si} \cdot LP, \qquad (1)$$

where $n_{Si}=3.3$ is the refractive index of silicon in the NIR and LP is the lattice parameter of the array. In this case (LP=590 nm), the Rayleigh anomaly is expected to occur at a wavelength of ca. 2 microns, in good agreement with the results of Fig. 5. We point out that the Rayleigh anomaly coincides with the onset of diffraction[30]. This means that for longer wavelengths than $\lambda_{RA}$, any feature occurring in the optical reflectance (or transmission) of the periodic metallic nanostructure can *solely* be due to excitation of non-propagating, localized plasmon-like polaritons.

Interestingly, the results of Fig. 5 also indicate that the *far-field* optical response of the inverted pyramid array is fairly insensitive to the degree of polarization of the incident light. All three curves (red, blue, green) of the Au-coated array fall essentially on top of each other, having only qualitatively insignificant differences. This is simply due to the perfect square symmetry of the pyramidal array and the mixed character of the plasmon polaritons, being exited in the NIR spectral range. As shown further below, locally the *near field* varies substantially depending on the polarization of the light with respect to the pyramids facets and corners.

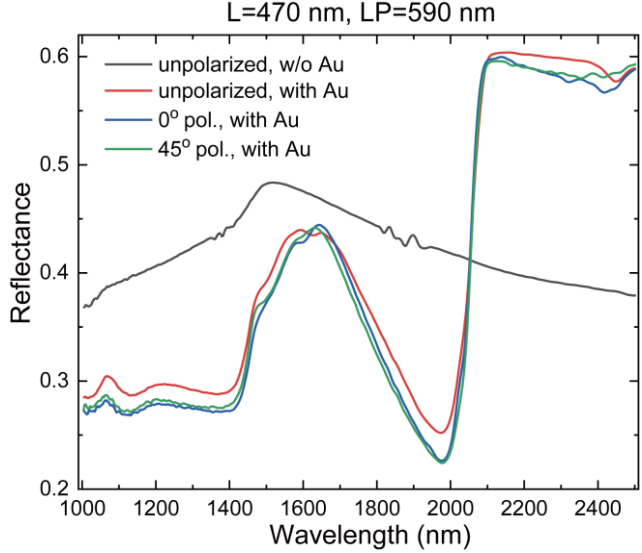

**Figure 5:** Measured reflectance spectra of an array of small inverted pyramids (L=470 nm, LP=590 nm) without and with Au layer for normal incidence from the Si substrate side and for different polarizations: unpolarised, 0º and 45º polarization.

The results for the optical response of the fabricated structures are summarized in Fig. 6, where we show the measured and calculated normalized absorbance in the NIR spectral range from 1000 to 2500 nm for two representative examples: An array with small (L=470 nm, LP=590 nm) and one with large (L=4.1 μm, LP=6.0 μm) inverted-pyramids covered or not with gold. The absorbance is directly obtained from the normalized reflectance (R) and transmittance (T), computed as A=1-R-T. We first point out that the agreement between experiment and theory is very satisfactory, qualitatively as well as quantitatively. In both cases, the introduction of the Au film leads to a strong increase in absorbance in the whole infrared region. For the small gold-covered inverted pyramids the overall measured absorbance is close to 80% for wavelengths below that of the Rayleigh anomaly. The two wavelengths, 1290 and 1950 nm, where the normalized absorbance reaches unity in the calculated response (see Fig. 6a), correspond to the excitation of strong plasmon-polariton resonances. The corresponding near-field distributions are discussed below. For the large pyramids, the addition of the Au layer also causes an increase of the absorbance, though the effect is moderate in comparison with the small pyramids. The average absorbance thus increases from ca. 25% to 40%. Since the transmittance is nearly zero, this means that much of the light in the NIR range is reflected by the large, microns-size pyramids (about 60% of the light in average). We point out that after Eq. (1) the Rayleigh anomaly is expected for the large pyramid array to occur at around 20 μm. This wavelength obviously falls outside the spectral range of our experiments. Hence, resonances associated with the various maxima observed in the spectra of Fig. 6b correspond to propagating Bragg photon-like polaritons. In general, for these large sizes and due to the very high refractive index of Si, the texturing of the Au/Si interface is not effective for IR light absorption, working the array as a reflection grating.

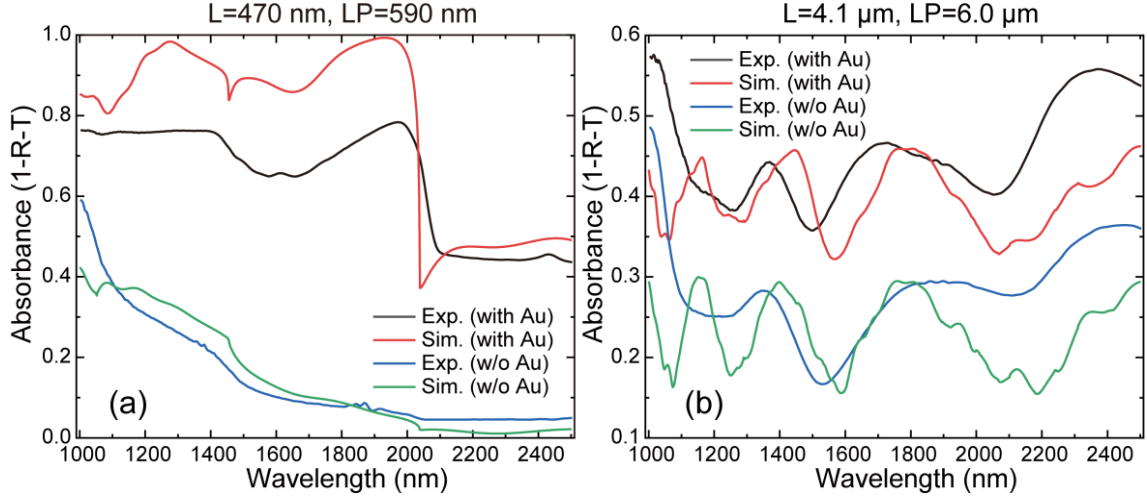

**Figure 6:** Measured and simulated absorbance (1-R-T) in the NIR spectral range of a representative **(a)** small (L = 470 nm, LP = 590 nm) and **(b)** large (L = 4.1 μm, LP = 6.0 μm) inverted-pyramid array with and without Au film.

Regarding the comparison between experiment and the numerical FDTD simulations, a clarification should be made. Whereas in the case of small pyramids the good agreement is almost for free, for large pyramids a special treatment of the theoretical data is required. We recall that the reflection experiments were carried out at normal incidence from the Si substrate side. Most of the light is then back-reflected from the textured Au/Si bottom interface. For large pyramids the onset of diffraction occurs at $\lambda_{RA}$ ~20 μm, thus, in the NIR range from 1 to 2.5 μm a substantial portion of the light will be diffracted into higher orders. The propagation direction of these high-index modes forms large angles with respect to the substrate normal. Due to the large refractive index of Si, most of these higher-order modes undergo total reflection at the top Si/air interface and, hence, they cannot be detected. In addition, the small numerical aperture of the microscope objective used in the experiments ends up further restricting the collection of light to the zeroth order. As a consequence, we have written a script to pick up *only* the zeroth order contribution to the reflectance out of the FDTD simulations. In this way, we are able to attain reasonably good agreement between theory and experiment for micron-size pyramids, for example, as shown in Fig. 6b. For submicron-size pyramids the zeroth order dominates the reflectance of the Au/Si inverted pyramid array, hence, no special treatment of the simulations is required.

In order to elucidate the potential of a given plasmon-polariton resonance for generating hot electrons by internal photoemission, we have analysed in detail the spatial distribution of the near electromagnetic field, obtained as the integrated response over the entire duration of the FDTD simulation. Calculations were carried out for certain special wavelengths using different planar monitors, corresponding to cross sections along pyramid facets (0º) or pyramid diagonals (45º) with light either S- or P-polarized (perpendicular or parallel, respectively) with respect to the corresponding monitor plane. Whereas the polarization turned out to be irrelevant for the far field, it is crucial for the local variations of the near field, the latter depending strongly on the plasmonic character of the polariton being excited.

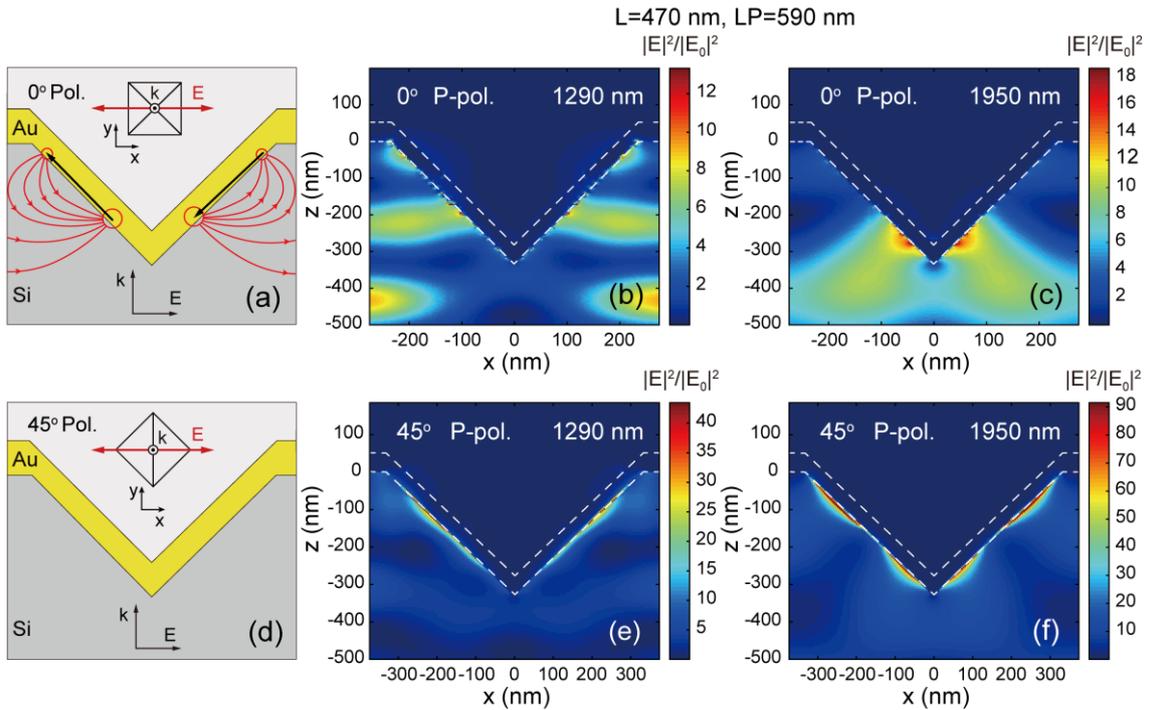

**Figure 7:** Near-field properties of small inverted pyramids (L=470 nm, LP= 590 nm) covered with 50 nm of Au. **(a)** Sketch of the geometry and polarization configuration for a planar monitor corresponding to a cross section through the middle of the pyramid along two facets (0º monitor). K is the wavevector of the light and E the electric field component. **(b)** and **(c)** Near-field distribution for P-polarized light in the 0º-monitor plane for resonances at 1290 nm and 1950 nm, respectively. **(d)** Idem (a) but for a monitor along diagonals of the pyramids (45º monitor). **(e)** and **(f)** Idem (b) and (c) for P-polarized light in the 45º-monitor plane.

Figure 7 shows the results of FDTD simulations as far as the near field is concerned, carried out for the same small pyramid array, whose far-field response was calculated previously, and for two special wavelengths, 1290 and 1950 nm, where the calculated reflectance reaches unity (see Fig. 6a). As illustrated by the contour plots of Fig. 7, for small pyramids the enhancement of the intensity of the electric-field component ($|E|^2/|E_0|^2$) for P-polarized light at the two selected wavelengths peaks mainly at different points in the very vicinity of the Au/Si interface. This is clear indication of the strong plasmonic character of these resonant modes, at least of its P-polarized component. Whereas for the 0º monitor the E-field intensity enhancement is typically greater than a factor ten, for the 45º monitor the enhancements can reach up to two orders of magnitude. In contrast, for S-polarized light (perpendicular to the plane of the monitors) the electromagnetic field is concentrated in regions close to the pyramid facets and tip but clearly separated from the Au/Si interfaces, as depicted in Fig. S4 of the ESI. This is evidence of an outspoken photonic character of the S-polarized component; the maximum intensity enhancement remaining always below one order of magnitude.

From the point of view of the possible application for the generation and injection of hot electrons into the semiconductor, it is clear that the S-polarized components are useless. In contrast, the near fields associated with the P-polarized components present two key features. On the one hand, they produce large concentrations of charges at points or localized regions right on the surface of the inverted pyramids, the so-called hot spots, which produce extremely high electric fields. On the other hand, the electric-field vectors are perpendicular to the metal surface at the hot spots. This can be visualized in Fig. S5 of the ESI, where we overlay the electric-field vectors to the intensity contour plots of Fig. 7 for the 0º monitor. An inspection of the field lines obtained from the FDTD calculations reveals that the hot spots on each facet correspond to the end-point charges of a sort of electrical dipole, oscillating with the same frequency of the incident light. Furthermore, the two dipoles in confronted facets are by symmetry in exact opposition. The sketch in Fig. 7a illustrates the disposition of both induced dipoles. The effective net charge at each end of the dipoles or the charge separation might vary for different modes (wavelengths) and different monitors but for pyramid of this size we always find that NIR light induces two dipoles in opposition on the pyramid facets. This has an important consequence for the spatial distribution of hot spots. In particular, this explains why, contrary to intuition, the pyramid tip is never a hot spot. The reason is that the electric fields generated by each but opposite dipole exactly cancel each other at the pyramid tip.

For large, microns-size pyramids their potential for hot-electron generation is less encouraging. Figure S6 of the ESI displays the contour plots of the electric-field intensity of an array of large pyramids (L=4.1 μm, LP=6.0 μm) at two resonant wavelengths of 1450 and 1760 nm for both P and S-polarized NIR light with respect to a 0º monitor. The field-intensity distribution reveals that these modes possess mainly photonic character, i.e. they are Bragg-like modes. Except for a few moderately intense hot spots for P-polarization at 1450 nm, all other field enhancements occur away from the metal surface in the space between the pyramids inside the silicon. The hot spots, however, now correspond to the end points of a series of dipoles aligned in head-to-tail manner along the pyramid facets.

Conclusions

In summary, an interesting alternative to current NIR light harvesters was presented, which is a plasmonic metamaterial based on micro/nanostructuring of silicon wafers by engraving arrays of inverted pyramids covered with a thin gold film, readily fabricated by scalable nanoimprint lithography combined with wet etching using KOH. For sub-micron-size pyramids the overall absorbance in the NIR spectral range (1 to 2.5 μm) can exceed 80%, which is shown to be due to the resonant excitation of plasmon polaritons sustained by the pyramidal array. Finite-differences time-domain simulations revealed that the near field associated with the plasmonic component of the polariton modes excited by NIR light presents several hot spots distributed throughout the Au/Si interface, characterized by strong enhancements of the electric-field intensity of up to two orders of magnitude. The existence of these hot spots is instrumental for the possible generation of photocurrents by internal photoemission of hot electrons from the metal into the semiconductor. In this way, it has been demonstrated the great potential of this type of plasmonic/photonic Au/Si nanostructures for their implementation not only in NIR harvesting devices but also for application in infrared photo-detection, photo-catalysis, or cell photo-stimulation in the IR biological window, for instance.

Acknowledgements

This project has received funding from the European Union's Horizon 2020 research and innovation programme under the Marie Skłodowska-Curie Individual Fellowship (MGA MSCA-IF) grant agreement No. 839402 (PLASMIONICO). The Spanish "Ministerio de Ciencia e Innovación (MICINN)" is gratefully acknowledged for its support through grants SEV-2015-0496 (FUNMAT) and CEX2019-000917-S (FUNFUTURE) in the framework of the Spanish Severo Ochoa Centre of Excellence program and the AEI/FEDER(UE) grant PGC2018-095411-B-100 (RAINBOW) and PID2019-106860GB-I00 (HIGHN). The authors also thank the Catalan agency AGAUR for grant 2017-SGR-00488. JLG-P acknowledges support of MSCA-IF fellowship grant agreement No. 840064 (2D_PHOT); JH acknowledges a fellowship from the China Scholarship Council (CSC) and the PhD program in Materials Science from Universitat Autònoma de Barcelona in which he is enrolled.

# Electronic Supplementary Information (ESI)

*Efficient infrared sunlight absorbers based on gold-covered, inverted silicon pyramid arrays*

Jinhui Hu[a], Luis A. Pérez[a,*], Juan Luis Garcia-Pomar[b], Agustín Mihi[a], Miquel Garriga[a], M. Isabel Alonso[a], Alejandro R. Goñi[a,c,**]

[*]lperez@icmab.es, [**]goni@icmab.es

[a]Institut de Ciència de Materials de Barcelona, ICMAB-CSIC, Campus UAB, 08193 Bellaterra, Spain

[b]INL-International Iberian Nanotechnology Laboratory, Av. Mestre José Veiga s/n, 4715-330 Braga, Portugal

[c]ICREA, Passeig Lluís Companys 23, 08010 Barcelona, Spain

Fabrication of inverted micro-pyramids:

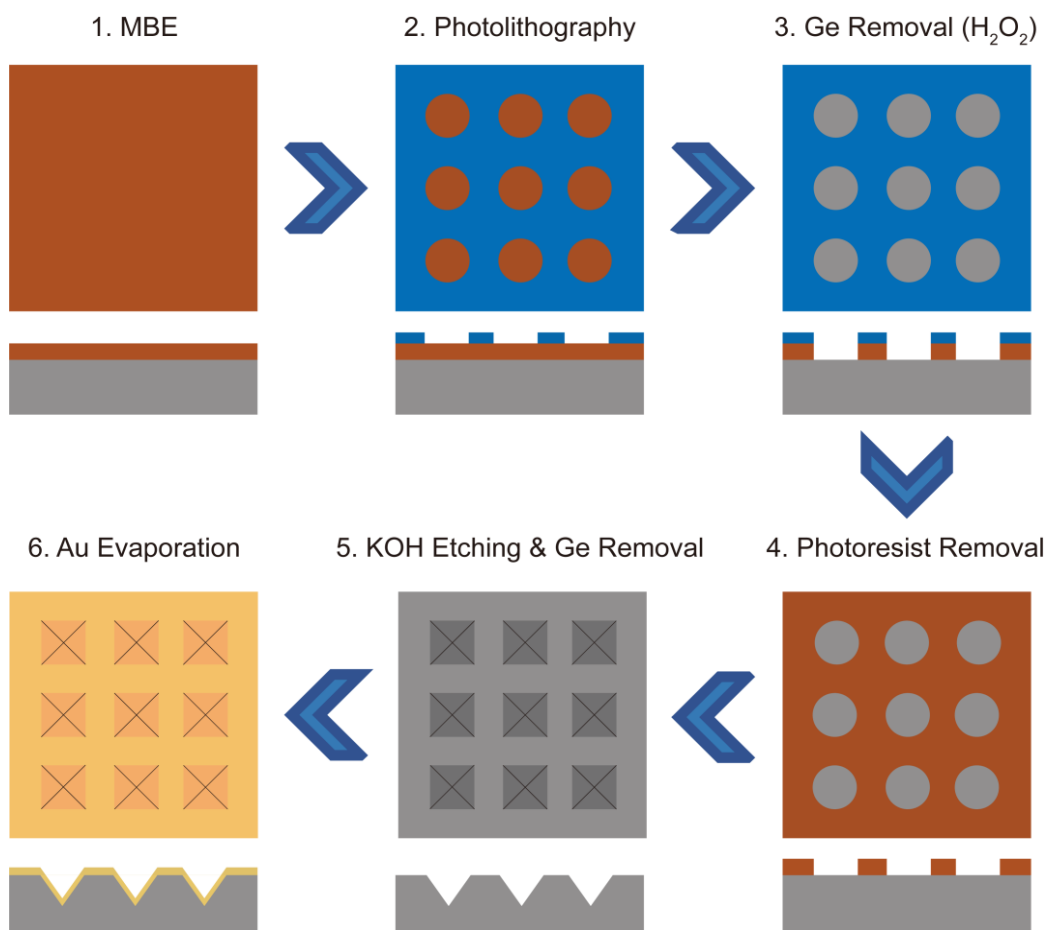

**Figure S1:** Different steps of the fabrication procedure of microns-size inverted pyramids by conventional photolithography and wet KOH etching.

A sketch of the different steps of the fabrication procedure by conventional photolithography and wet KOH etching is displayed in Fig. S1. For the fabrication of microns-size pyramids a 1200 nm-thick Shipley resist was deposited on Si/Ge heterostructure. Then, a pattern of holes with the desired diameter and pitch is printed on the resist with a microwriter under exposition with a UV laser, subsequently developed and rinsed with water. A wet $H_2O_2$ etching was carried out to remove rest of the Ge in the holes. An anisotropic wet etching in 33wt% KOH produced the inverted pyramids in the exposed Si. Finally, a 50 nm thick Au film was deposited by thermal evaporation on the surface of inverted pyramid.

*Numerical Simulation*:

As shown in Fig. S2, the unit cell of the inverted pyramid structure was set as the simulation region using periodic boundary conditions (PBC) along the x and y axes and perfectly matched layer conditions along the z-axis. The pyramid and the surrounding medium were divided into a mesh (not shown) for computing the finite differences.

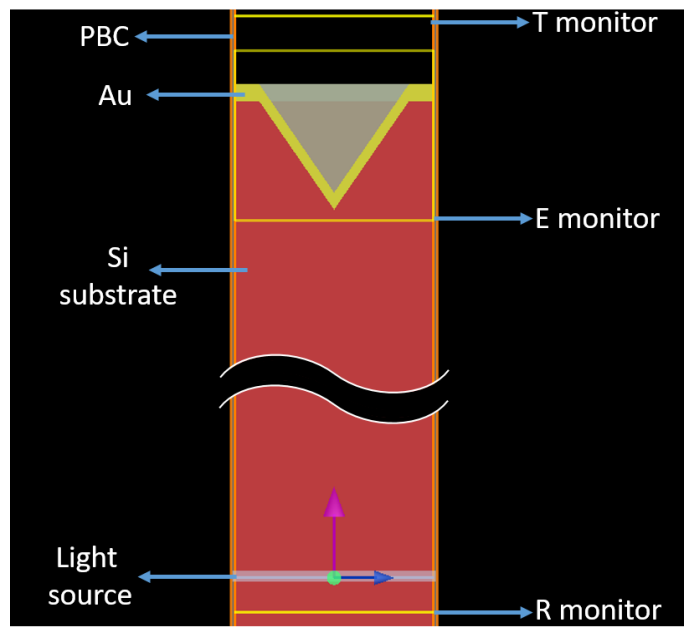

**Figure S2:** FDTD simulation model of the inverted-pyramid unit cell with periodic boundary conditions (PBC). Different elements are indicated by arrows like the light source, the Si substrate, the gold layer as well as the reflection (R), transmission (T) and electric-field (E) monitors.

*Pre-screening of pyramid array parameters*:

In Fig. S3, we show the pre-screening of the gold layer thickness. The result is straightforward: With increasing Au thickness light absorption increases rapidly from zero to a saturation value, reached above a thickness of ca. 30 nm. Consequently, we deposited routinely 50 nm of gold, so as to obtain smooth, strongly-absorbing Au layers.

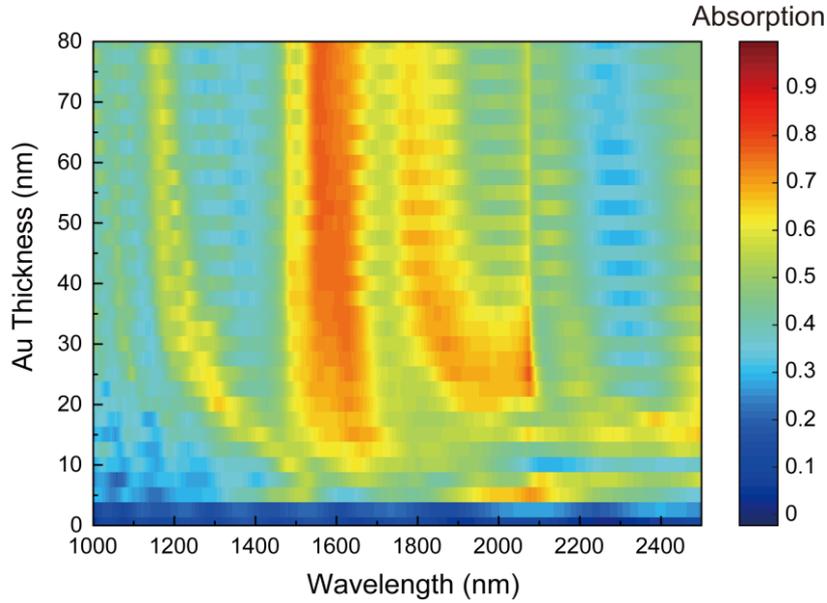

**Figure S3:** Contour plots of the normalized absorption as a function of the Au film thickness for a structure with L= 500 nm, LP= 600 nm.

*Calculated near-field properties*:

For S-polarized light (perpendicular to the plane of the monitors) the electromagnetic field is concentrated in regions close to the pyramid facets and tip but clearly separated from the Au/Si interfaces, as depicted in Fig. S4. This is evidence of an outspoken photonic character of the S-polarized component of both resonant modes. In this case, the maximum intensity enhancement remains always below one order of magnitude.

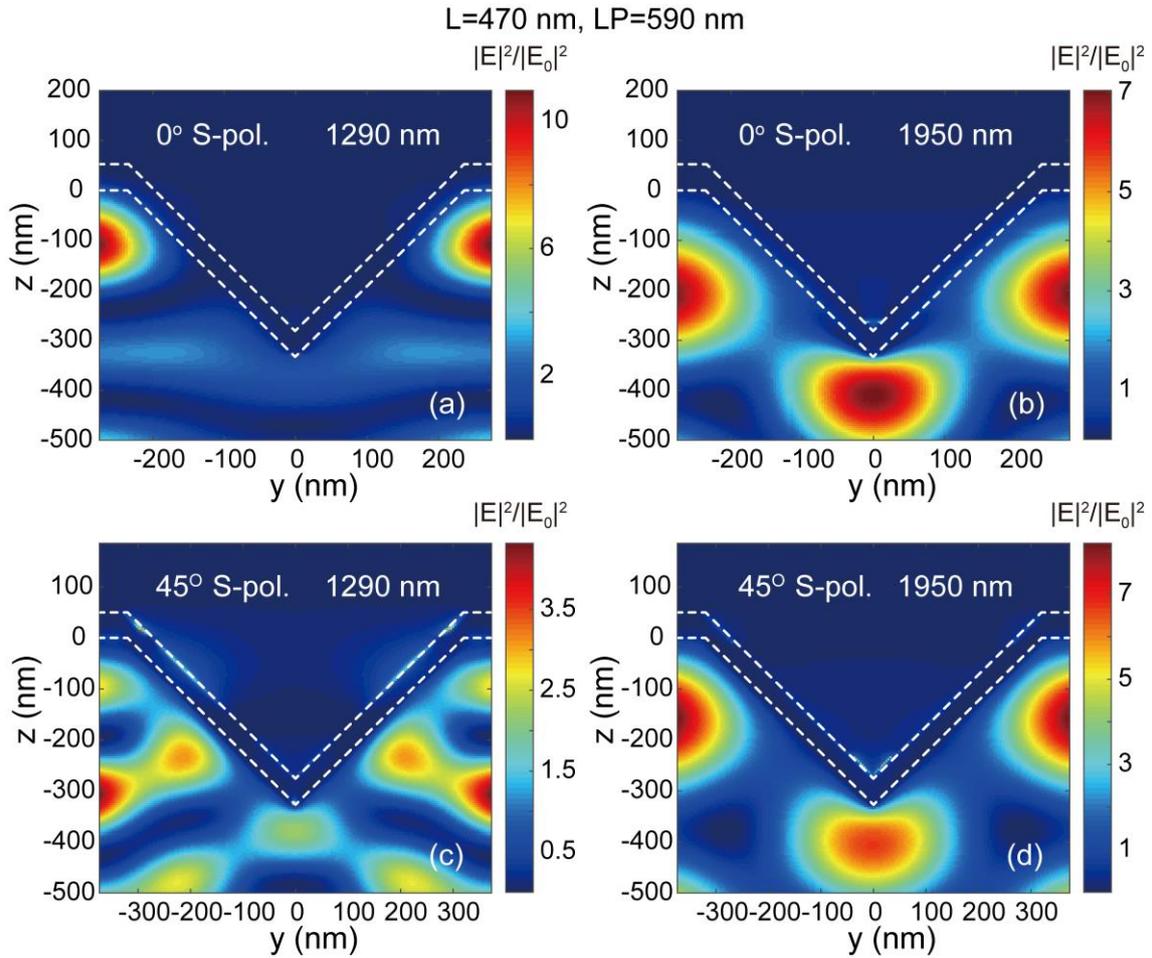

**Figure S4:** Near-field properties of small inverted pyramids (L=470 nm, LP= 590 nm) covered with 50 nm of Au. The light is in all cases S-polarized, i.e. perpendicular to the monitor plane. **(a),(b)** Case of 0° monitor, S-polarized light of 1290 nm and 1950 nm, respectively. **(c),(d)** Case of 45° monitor, S-polarized light of 1290 nm and 1950 nm, respectively.

In contrast to the S-polarized components, the near fields associated with the P-polarized components produce large concentrations of charges at points or localized regions right on the surface of the inverted pyramids, the so-called hot spots, for which the electric-field vectors are perpendicular to the metal surface at the hot spots. This can be visualized in Fig. S5, where we overlay the electric-field vectors to the intensity contour plots of Fig. 7 of the manuscript for the 0° monitor. An inspection of the field lines obtained from the FDTD calculations reveals that the hot spots on each facet correspond to the end-point charges of a sort of electrical dipole, oscillating with the same frequency of the incident light. Furthermore, the two dipoles in confronted facets are by symmetry in exact opposition. The effective net charge at each end of the dipoles or the charge separation might vary for different modes (wavelengths) and different monitors but for pyramid of this size we always find that NIR light induces two dipoles in opposition on the pyramid facets.

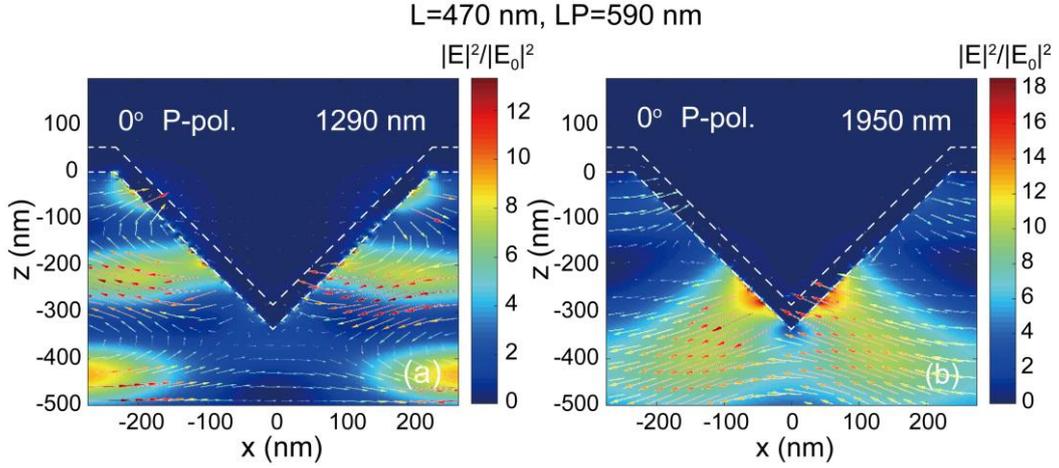

**Figure S5:** Visualization of the electric-field vectors for the near-field distributions calculated for P-polarized light, 0º monitor, at **(a)** 1290 nm and **(b)** 1950 nm for a small Au-covered, inverted pyramid array (L=470 nm, LP= 590 nm), corresponding to Figs. 9b and 9c.

For large, microns-size pyramids their potential for hot-electron generation is less encouraging. Figure S6 displays the contour plots of the electric-field intensity of an array of large pyramids (L=4.1 μm, LP=6.0 μm) at two resonant wavelengths of 1450 and 1760 nm for both P and S-polarized NIR light with respect to a 0º monitor. The field-intensity distribution reveals that these modes possess mainly photonic character, i.e. they are Bragg-like modes. Except for a few moderately intense hot spots for P-polarization at 1450 nm, all other field enhancements occur away from the metal surface in the space between the pyramids inside the silicon. The hot spots, however, now correspond to the end points of a series of dipoles aligned in head-to-tail manner along the pyramid facets.

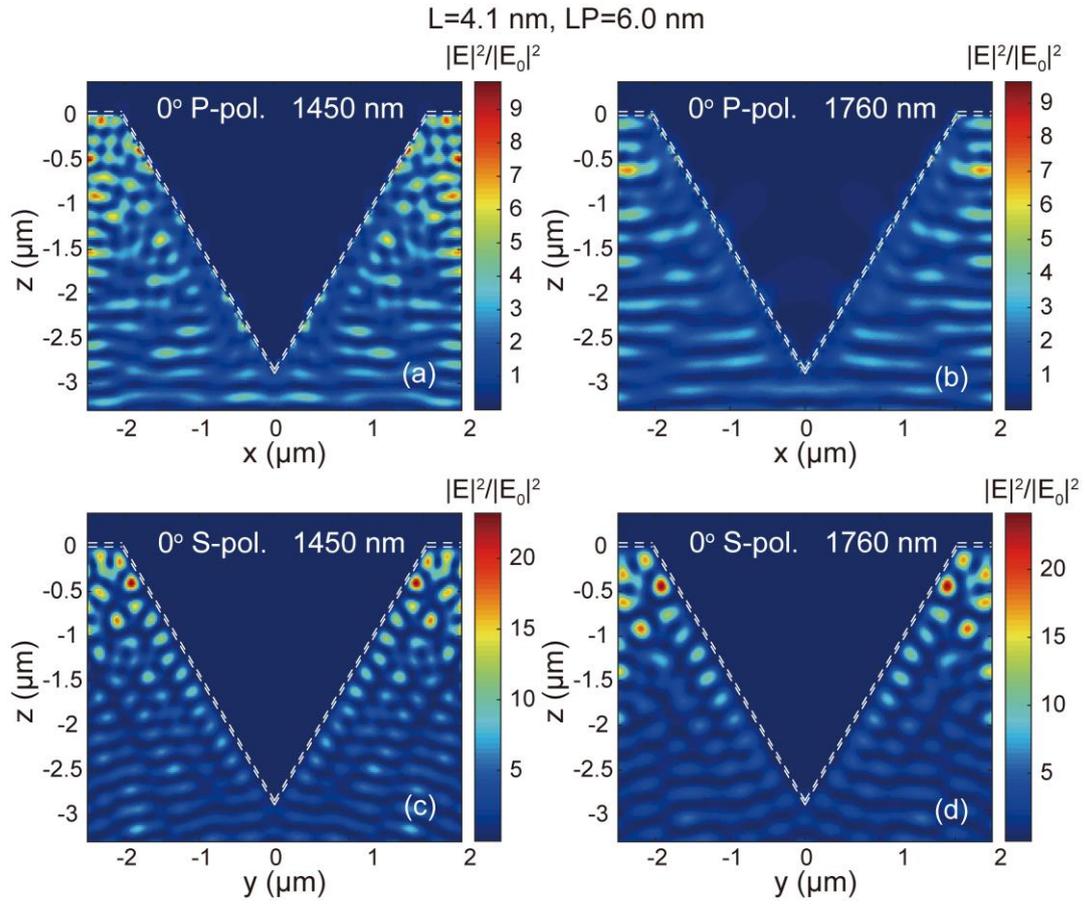

**Figure S6:** Near-field distribution of the electric-field intensity across 0° monitors for **(a), (b)** P-polarized light of 1450 nm and 1760 nm, respectively, and **(c), (d)** for S-polarized light of 1450 nm and 1760 nm, respectively, in the case of an array of gold-covered, large inverted pyramids (L=4.1 μm, LP= 6.0 μm).